\documentclass[11pt,preprint,flushrt]{aastex}
\usepackage{amsmath}
\usepackage{times}
\allowdisplaybreaks[1]

\newcommand\etal{{\it et al.\/}}
\newcommand\ie{{\it i.e.\/}}

\begin{document}

\title{Principal Component Analysis of PSF Variation
in Weak Lensing Surveys}

\author{Mike Jarvis, Bhuvnesh Jain}
\affil{Dept. of Physics and Astronomy, University of Pennsylvania,
Philadelphia, PA 19104}
\email{mjarvis, bjain@physics.upenn.edu}

\begin{abstract}

We introduce a new algorithm for interpolating measurements of 
the point-spread function (PSF) using stars from
many exposures.  The principal components of the 
variation in the PSF pattern from multiple exposures are used
to solve for better fits for each individual exposure. 
These improved fits are then used to correct the weak lensing shapes.  
Since we expect some degree of correlation in the PSF anisotropy
across exposures, using information from stars in all exposures 
leads to a significant gain in the accuracy 
of the estimated PSF. It means that in general, the accuracy of PSF
reconstruction is limited not by the number density of stars per
exposure, but by the stacked number density across all exposures 
in a given survey.  This technique is applied to the 75 square degree 
CTIO lensing survey, and we find 
that the PSF variation is well described by a small number of 
components. There is a significant improvement in
the lensing measurements: the residual stellar PSF correlations are reduced
by several orders of magnitude, and the measured B-mode in the two-point
correlation is consistent with zero down to 1 arcminute.
We also discuss the applications of the PCA technique to future surveys. 
\end{abstract}


\section{Introduction}

There are quite a few different systematic errors which can contaminate
a weak lensing signal at the level of typical cosmic shear measurements.
The largest of these has been the corrections of the anisotropic
point-spread-function (PSF).  Stars, which probe the PSF, are seen to have
significantly elliptical shapes.  Furthermore, on large cameras, the PSF
can vary substantially across the image.  Since the galaxy shapes are
convolved by the PSF before you observe them, the effects of the PSF
must be removed before performing any weak lensing analysis.

A number of techniques for removing the effects of the PSF have been
proposed \citep{KSB,Ho98,Ka00,BJ02,RB03}.  Recent
lensing meausrements \citep{vW02,Ho02,Re02,Ba03,Br03,Ha03,Ja03} 
have implemented these techniques well enough that the systematic
errors are generally somewhat smaller than the statistical errors.
However, as the size of weak lensing surveys increases and the statistical 
errors keep going down, it becomes more important to similarly reduce 
the systematic errors.  

This paper addresses the particular
difficulty of interpolating the PSF between the stars in the image. 
Typically, the procedure has been
to fit the various components of the PSF shape (the ellipticity, size,
and possibly higher order components) to a polynomial function across the
image.  This technique is limited to probing variations in the 
PSF pattern which vary more slowly than the typical spacing between
the stars.  In fact it is somewhat worse than this, since
the stellar shape measurements are often noisy, and there may be 
small galaxies masquerading as stars which have shapes which do not
represent the PSF.  Therefore, a smoother function with 
outlier rejection is required.

To address this problem, \citet{Ho04}
uses dense stellar fields to measure the high-order component of the 
PSF pattern, and then uses the stars in the individual images for a
low-order correction to this pattern.  This works well when the 
PSF pattern is reasonably stable so that the dense fields give a 
good first estimate of the pattern.  However, in general, and for 
our data in particular (see \S\ref{pcasection}), this will not
be the case.
\citet{vW04} get better interpolation results when they switch from
polynomials to rational functions.  Rational functions are almost
always better at modeling complicated functions, so 
they do a better job of modeling the PSF variation.

However, both of these methods do leave some residual B-mode contamination
(see \S\ref{statssection} for a discussion of the B-mode), indicating
that there is still some uncorrected PSF anisotropy in the data.
The residual is predominantly at small scales where high-order 
wiggles in the PSF pattern are not well modeled.  
In fact, it has been suggested that this level of B-mode may be an
irreducible systematic for ground-based cosmic shear measurements,
since there are simply too few stars to model the real PSF variation
on these scales.

While it is likely true that there is some irreducible systematic
effect on small scales, we show in this paper that it is much smaller
than the previous results have indicated.  The PSF pattern is found to be 
largely a function of only a few variables.  Therefore, we are able to
improve the fits of this pattern by using stars from many exposures.
The limits to how well the PSF can be modeled are therefore governed not
by the number of stars in a single exposure, but rather by the total number
of stars from all of the exposures available.  

In \S\ref{datasection} we briefly review our CTIO survey data, largely
deferring to \citet{Ja03} for more details.  We describe our new
PSF modeling algorithm in \S\ref{pcasection} and 
conclude in \S\ref{discussionsection}.

\section{Data}
\label{datasection}

Our CTIO survey data was originally described in detail in \citet{Ja03},
and we refer the reader to that paper for the details about the
data and the analysis.  Here, we present merely a brief summary 
of the data set, and point out two changes from our previous analysis - 
the dilution correction and the PSF interpolation.

We observed 12 fields, well separated
on the sky, each approximately 2.5 degrees on a side, giving us a total of 
75 square degrees.  We have useful galaxies for weak lensing in the magntude
range $19 < R < 23$, which gives about 2 million 
galaxies to use for our lensing statistics.

The shape measurements of the galaxies follow the techniques of 
\citet{BJ02} using the convolution kernel method described therein.
The first change from our previous analysis involves the dilution 
correction.  \citet{Hi03} point out that our formula for the dilution 
was incorrect with respect to the kurtosis of the PSF.  For the present
analysis, we have incorporated their linear PSF correction
which is described in Appendix B of their paper.  

The second change we made is in the PSF interpolation.  
Since the convolution kernel is only measured where we have an observed
star, and the PSF is far from uniform across each image, 
we need to interpolate between the stars.  In our previous analysis,
we used a separate fit for every image.  For this analysis, however,
we use our new PCA technique which is described in \S\ref{pcasection}.

Finally, we estimate the shear, $\gamma$, from an ensemble of ellipticities
using the ``easy'' weighting given in \citet{BJ02}, equation 5.36, and 
using the corresponding responsivity formulae. 

\section{PCA Analysis of PSF fitting}
\label{pcasection}

It was clear from our previous analysis that
the pattern of the PSF variation is not generally a simple function.
In our previous analysis, we had found that fourth order polynomials were
insufficient to fully describe the pattern.  Unfortunately, with only a
hundred or so stars per image, many of which are noisy and/or outliers,
the data did not well constrain fits to higher order than this.  We had
slightly improved upon this by using a smoothing spline function, but
that also failed to fully describe the variation.

This section describes our new technique of using Principal Component
Analysis (PCA) to allow us to use all of the stars in every image to 
make a global fit of the PSF variation.  This allows us to fit to a higher
order polynomial in $x$ and $y$, 
which seems to describe essentially all of the 
real variation of the PSF.

\begin{figure}[t]
\epsscale{1.0}
\plotone{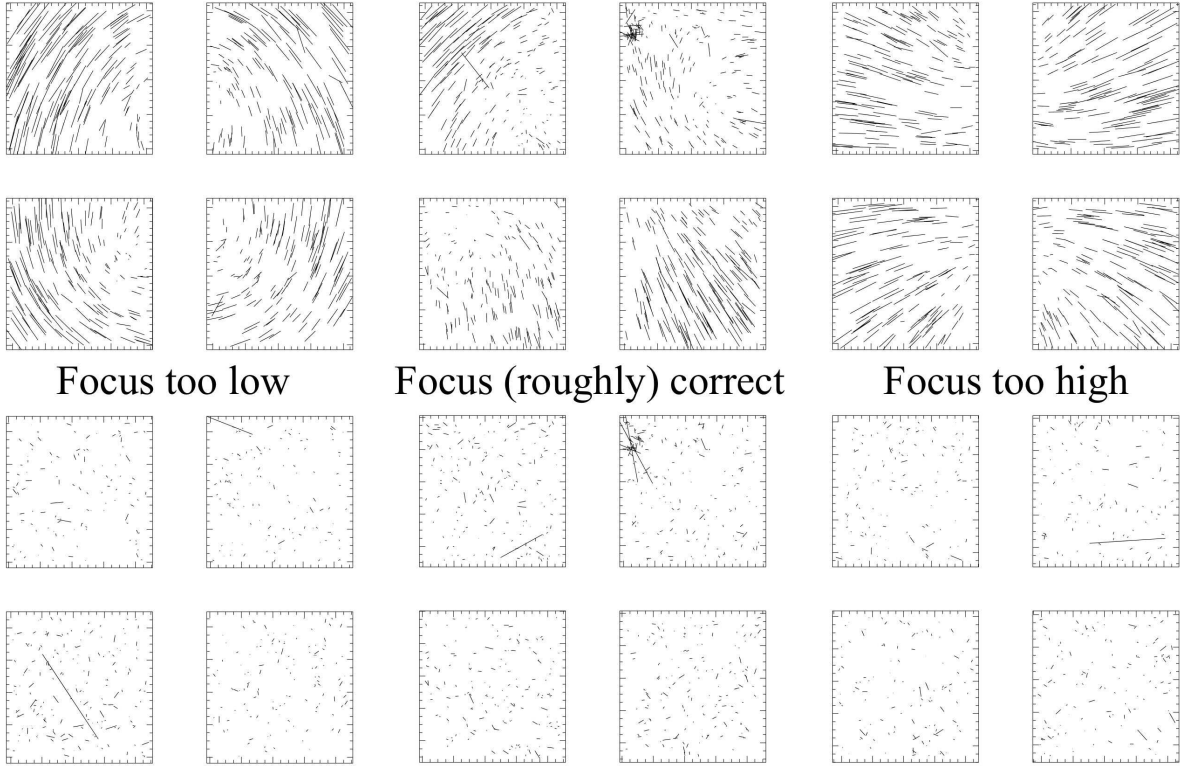}
\caption[]{\small
Whisker plots of the stars for three of our BTC exposures.  The four
squares per exposure correspond to the four camera chips, each $15'$ on
a side. 
In all of the plots, the lengths of the lines
are proportional to the ellipticities of the stars, and the orientation is
in the same direction as the direction of the ellipticity.
The top row show the shapes from the raw images, and the bottom row
shows the shapes of the same stars after correction using the PCA technique.
The left and right plots represent the two extremes in defocus.
The middle is more typical of most of our data. 
}
\label{whiskerplot}
\end{figure}

Figure~\ref{whiskerplot} shows before and after ``whisker plots'' 
for three of our 
exposures.  These exposures were taken with the Big Throughput Camera (BTC),
for which the PSF patterns were significantly worse than with the Mosaic
camera.\footnote{
Our entire survey includes approximately equal area of coverage for 
the two cameras.
} Each line (``whisker'') represents the ellipticity of an observed star.  The 
length of the line is proportional to the ellipticity, and the 
orientation is the same as that of the star.
The top row shows the plots for the three exposures before any corrections.
The leftmost plot represents the most out of focus image in one direction,
and the rightmost represents the extreme in the other direction.\footnote{
Specifically, these images have the largest and smallest  
$U(i,1)$ value, which we describe in \S\ref{methodsection}.}
The middle plot is far more typical, but apparently the four chips are 
not precisely coplanar, so they do not all come into focus at exactly the 
same time.  The bottom row are the stars from the same three images after 
our corrections.  

We note that observing a dense stellar field, as proposed by \citet{Ho04},
is not sufficient for our survey, since the PSF pattern 
changes dramatically with the relative focus position.  Using the high order 
fit from some ``average'' image will not help any image with a 
different focus value.  

\subsection{PCA Method}
\label{methodsection}

The first step in our PCA technique is to do a polynomial
fit of the PSF variation in each 
image.\footnote{Technically, for our analysis, we fit the kernel components,
rather than the PSF shape, but as most other lensing analysis techniques 
do not use reconvolution, we note that the method works exactly the same
when fitting the PSF shapes directly.}
This should be done to a high enough order that the fit describes the
bulk properties of the pattern well, but not so high that it gets pulled
by noisy measurements.  For our data we found that a fourth order fit
was adequate. 

We arrange these numbers into a matrix $M$ where each row of the matrix 
corresponds to a different exposure, and the elements in the row are 
the numbers from the above fit.  Thus for $N_{\rm exp}$ exposures
and a fourth order polynomial for each exposure, $M$ is an 
$N_{\rm exp} \times 15$ matrix, since there are 15 coefficients in a 
fourth order polynomial of $x$,$y$.
We then perform a singular value decomposition of this matrix:
\begin{equation}
M = U S V
\end{equation}
where $U$ and $V$ are unitary matrices, and $S$ is diagonal.  
The rows of $V$ are called the principal components, and the elements of 
$S$ are called the singular values.  These are in descending order,
so that $S(1,1)$ is the largest singular value.

Each row of $M$ can now be written as a weighted sum of the principal
components:
\begin{equation}
M(i,*) = \sum_k U(i,k) S(k,k) V(k,*), 
\end{equation}
where the index $i$ denotes the exposure number. 
Most of the variation in the rows of $M$ is described by the first principal 
component ($k = 1$).  For our data, since focus seems to be the dominant
cause of the variation in our PSF pattern, this coefficient is basically 
telling us the focus position of that exposure in some arbitrary metric. 
So we can think of $U(i,1)$ as the focus position. 
While the first principal component is indeed fairly dominant,
the next several components are not insignificant.  The magnitude
of each singular value $S(k,k)$ indicates the significance of that
component, so one should use the components up to some $k_{\rm max}$ 
where the singular values become suitably small.

Thus for each star we now have its position, given by $x$, $y$, 
and the above $U(i,k)$ for its exposure.  
We can now find a global best fit function of these values
using the stars in all of our exposures.  
This will give us a better fit than we had fitting each image individually.
The form of this fitting function is somewhat arbitrary.  We choose to
make a fit which is polynomial in $x$ and $y$ and linear in $U(i,k)$:
\begin{equation}
e_{\rm psf}(x,y,i) = \sum_k U(i,k) P_k^{(n)}(x,y)
\label{fiteqn}
\end{equation}
where $P_k$ are $n$th order polynomial functions.
We do not need to consider higher order terms in $U(i,k)$, since if 
the PSF varied as some function of $U(i,1)$ (for example), 
instead of linearly, then the 
singular value decomposition would have picked those values out as the 
$U(i,1)$ instead. 
We also do not include a term with no $U(i,k)$ factor, since
to the extent that such a term
is important, it will be represented as a linear combination of some of 
the $U(i,k)$ values.

The real advantage to this fit is that we can now use all of the stars from 
every image to constrain the fit parameters.  There are $k_{\rm max}$
times as many coefficients to
constrain, but we have $N_{\rm exp}$ times as many stars to do the 
constraining.  ($N_{\rm exp}$ will generally be at least an
order of magnitude larger than $k_{\rm max}$.)

In addition to providing a better fit in general, it also allows us 
to take $n$ (the order of the polynomial fits) to be significantly higher than 
we were able when we were fitting one image at a time.
Another advantage is that this technique allows for better outlier rejection,
since there is less worry about rejecting all of the stars in a region 
where the PSF changes quickly, and with more stars to consider, the outliers
are more obvious.

Obviously, there are some choices one can make about how many principal
components to use (\ie\ what is $k_{\rm max}$?) and what order to use
for the second set of fits (\ie\ what is $n$?).  For our analysis, we
used up to the principal component whose singular value was 0.01
times the largest singular value, which corresponded to $k_{\rm max}$ of 
around 30 for our data.  For the BTC data, we chose $n=10$, and for the
Mosaic data, $n=8$, since the Mosaic PSF patterns
were smoother and did not require quite as high a fitting order. 

We should point out that using values of $k_{\rm max}$ and $n$ as high as these
is computationally quite demanding.  There are $k_{\rm max}$ nth order 
polynomials, which have a total of $k_{\rm max} (n+1)(n+2)/2$ coefficients.
For our BTC fits, this number is around 2,000.  
Furthermore, we have 651 BTC exposures, which have a total of almost
100,000 star measurments for each chip.  So a least squares fit
involves a QR decomposition of a $100,000 \times 2,000$ matrix, which takes 
of order a day to compute on a desktop computer\footnote{
Explicit construction of the design matrix takes a similar amount of time.}.
The computing time
for this step scales as $N_* k_{\rm max}^2 n^4$, where $N_*$ is the 
total number of star measurements.

\subsection{Prospects for Future Surveys}

For future surveys which plan to observe many times the area of the 
CTIO survey, the number of stars which can be used to constrain the PSF
pattern will increase correspondingly.  Thus, to the extent that the 
pattern is really a function of only about $k_{\rm max}$ parameters, the 
systematic errors from the PSF fitting will go down as $\sqrt{N_{\rm exp}}$ 
along with the statistical errors.

Furthermore, the fits are only accurate down to a scale of $1/n$ times 
the chip size.  For the CTIO data, tenth order fits can correct the
correlations down to about $1\arcmin$.  To accurately correct for the 
PSF down to $0.5\arcmin$ requires doubling $n$.  With more stars, 
future surveys will have the ability to constrain higher order fits 
such as this. However, since the computing time for the
fits scales as $n^4$, such a calculation would take 
about 16 times as long. So correcting scales much smaller than this would 
require the use of substantial computing resources. 

Note that the description of the $U(i,k)$ values as ``focus'' values
should not be construed to mean that only focus is important.  
The PCA technique automatically picks out the important effects on the PSF
pattern, regardless of the cause, and corrects for them.

For example, ground based telescopes also have systematic errors due to 
gravity, thermal effects and various other instrumental effects (see 
Section 4 for a discussion of atmospheric effects). 
These are likely to have some
degree of correlation across exposures, and are generally described 
by a small number of variables -- certainly much smaller than the 
number of exposures expected in planned surveys. 

In space, the principal variation in the PSF will most likely be due
to other factors than the focus, since the focus is more stable than for 
ground-based telescopes.  However, even in space the PSF pattern can vary
due to effects such as the sun's relative position, the position in 
the orbit, periodic mechanical and thermal adjustments or other factors. 

One potential improvement to this technique would be to incorporate the 
idea of \citet{vW04} to use rational functions rather than
polynomials.  They found that rational functions describe the PSF 
variation better than polynomials, so presumably that would still be the
case using our PCA technique.  The reason we did not switch to rational
functions is that they require a nonlinear fit to determine
the parameters.  Since we already have very large matrix equations which
take quite a while to solve, we did not want to make the calculation
take even longer by introducing a nonlinear fit.  Also, as our results with
polynomials are already consistent with no B-mode, it would be difficult
to test for any further improvement.  However, we do expect 
that this change could improve the quality of the fits even further. 

\subsection{Tests of Effectiveness}
\label{statssection}

The bottom row of Figure~\ref{whiskerplot} show the ellipticities of 
the stars for three of our BTC images after application of our PCA
corrections.
The whiskers are not precisely zero, but are consistent
with measurement noise.  In particular, there are no longer any 
apparent correlations between the whiskers.

\begin{figure}[t]
\epsscale{0.5}
\plotone{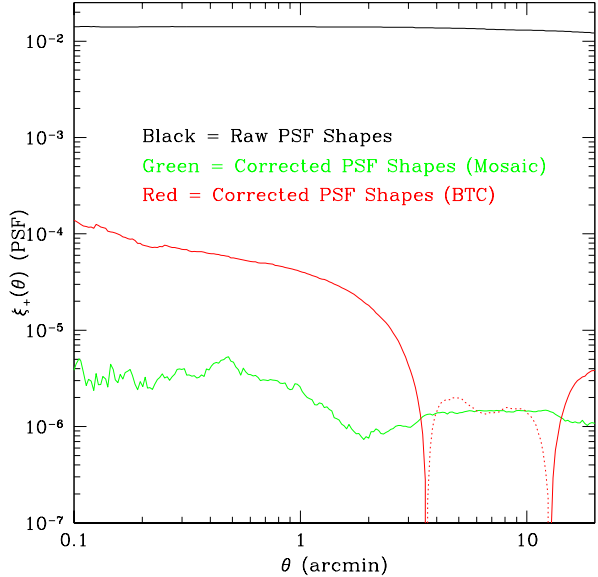}
\caption[]{ \small
The two-point correlation function of the PSF measurements.  The upper
black curve shows the raw, uncorrected star shapes.  The green and 
red curves
show the corrected shapes from the Mosaic and BTC images respectively.
The dotted portion of the BTC curve indicates negative values.
}
\label{starcorr}
\end{figure}

To better test this result, we can measure the two-point correlation function,
$\xi_+(\theta)$,
for all of the corrected stars in all of the images, limiting to
pairs of stars which were observed in the same image.  The results of
this test are in Figure~\ref{starcorr}, along with
the uncorrected correlation function\footnote{
Technically, the plotted curve for the raw shapes is for the Mosaic
data.  The raw BTC curve is very similar, but slightly higher.}.
For the Mosaic data, the power in the correlation function dropped 
by almost four orders
of magnitude over the entire range from $0.1\arcmin$ to $15\arcmin$ 
(the size of the chips).  
This is remarkable, since the eighth order polynomials were
only expected to work well down to about $1\arcmin$.
This corresponds to a factor of almost 100 reduction in the 
systematic shape of each star.  Furthermore, the effect of this 
residual on galaxy shapes is smaller still by roughly the ratio of
the PSF and galaxy sizes, which is typically about $1/2$ on average.

The results for the BTC data are similar above $3\arcmin$, but are
somewhat worse below this, where the tenth order fits were apparently
insufficient to model the actual patterns.
The ringing at higher scales 
is also indicative that a polynomial is not a very good description of the 
PSF patterns.  It is possible that a different functional
form would give better results for these data.  However, we note
that cameras for future surveys will almost certainly not be as warped
as the BTC camera, so the success of the PCA algorithm with the
Mosaic data is more relevant for upcoming surveys.
For cosmology constraints, we generally limit our consideration
to larger scales on which the PSF is well corrected, so the relatively
poor corrections at $\theta < 1\arcmin$ do not affect the results.

\begin{figure}[t]
\epsscale{0.5}
\plotone{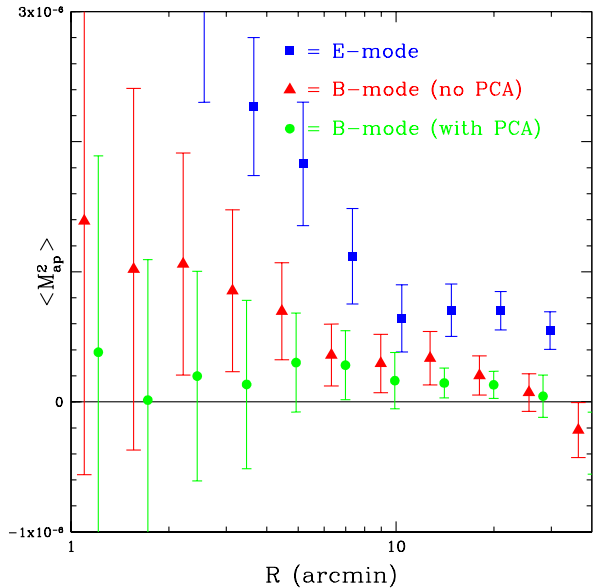}
\caption[]{ \small
The B-mode aperture mass measurements for our CTIO survey data.
The triangles and circles are the B-mode with and without using the PCA
technique, respectively.
The squares are the E-mode measurements.
}
\label{mapfigure}
\end{figure}

As another test, we can measure the divergence-free B-mode 
of the shear field for the corrected galaxies.
Since weak lensing should only produce a curl-free E-mode signal, 
a non-zero B-mode indicates the presence of some
residual systematic error.
One particular statistic which can measure the B-mode is 
the aperture mass statistic \citep{Sch98,Cr02,Sch02,Pen02},
which has been used quite oftern for cosmic shear studies.
Every weak lensing study which has tested 
for B-mode has detected some, although usually at a level somewhat less
than the E-mode signal and concentrated at smaller scales.  

Figure~\ref{mapfigure} shows the results for our reanalysis\footnote{
Note that $R$ in Figure~\ref{mapfigure} is not the same as $\theta$
in Figure~\ref{starcorr}.  $R = 1\arcmin$ corresponds roughly to 
$\theta = 2\arcmin$.}.
The triangles are the B-mode for fits on a chip-by-chip basis.
The circles are the B-mode for fits using the PCA technique.
We also plot the E-mode as squares for reference.
(The B-mode with no corrections would be off the top of the chart.)
The B-mode is now seen to be consistent with zero at nearly all scales,
which was not the case for our previous analysis. 
Our constraints on cosmology using these statistics are presented in 
\citet{JJB05}.

\section{Discussion}
\label{discussionsection}
We have presented a principal component approach to fitting 
the PSF anisotropy in imaging data. The stellar PSFs measured
in different pointings are combined to obtain a precise estimate
of the PSF at the position of each galaxy image.
Applying the technique to 
the CTIO survey data previously presented in \citet{Ja03},
we find that the measured lensing signal has
a much lower systematic contamination. We show two tests: the
ellipticity correlation of stars after the PSF correction, and
the residual B-mode in the two point correlation of galaxies. 

There has been significant recent progess in lensing analysis
in all of the critical steps leading to shear estimators: 
the measurement of galaxy shapes \citep{Ka00,BJ02,Re03,Ma04},
the deconvolution of the PSF to correct the measured shapes 
\citep{Ka00,BJ02,RB03}, 
and the estimation of the PSF at galaxy positions using stellar 
PSFs (\citealp{vW04}; this work).
These developments are important if future lensing
surveys are to achieve an accuracy in measurements that is 
close to the statistical errors. However, there is more work to be done
in actually implementing and testing all the pieces that have
been presented in the studies cited above. 

The principal component approach is applicable to any survey 
data, since it does not rely on any particular model for the
origin of the PSF shape. It simply finds close to the optimal
way of using all stellar PSF measurements. 
We believe it is well suited for the analysis of planned
lensing surveys that will cover greater area than ours, such
as the Dark Energy\footnote{http://cosmology.astro.uiuc.edu/DES/},
PanSTARRS\footnote{http://pan-starrs.ifa.hawaii.edu/},
LSST\footnote{http://www.lsst.org/}, and SNAP\footnote{http://snap.lbl.gov/} 
surveys. These surveys
will have many more pointings than ours, which will enable
more accurate fitting of the PSF, since the stacked density of
stars is a good indicator of how well the PCA technique can fit the
PSF. Statistical errors decrease with increasing survey area, and 
now these surveys can improve their systematic
corrections with increasing survey area as well.  

We note that if the atmosphere contributes truly random variations with 
significant power at high spatial frequency, then our technique will not be 
able to fit for this contribution to the PSF shape.  
However, such a component is expected
to be suppressed for typical weak lensing observations which 
are taken over 5 minutes or longer.  
\citet{KTL00} estimate the coherence time for the atmosphere's
contribution to the PSF to be of order a second. 
With exposures which are hundreds of times longer 
than this, any anisotropic component of the PSF shape from the atmosphere
will be highly circularized.  

Furthermore, a component of PSF variation that is truly random across 
exposures is easy to correct with an appropriate survey strategy. 
In calculating the shear correlation function (from which statistics 
like the aperture mass described above are derived),
one can use only pairs of galaxies that come from different exposures.  Then
the systematic correlation due to the random component of the PSF variation
cannot contaminate the signal. On scales
larger than the field of view, this happens automatically. 
For smaller scales, one needs to observe each location multiple times.
This is already common practice to help remove cosmic rays and spurious
detections.  Using this procedure, the statistical errors
would increase by about $\sqrt{N/(N-1)}$ where $N$
is the number of exposures per location, but with modest $N$, 
this is a small price to pay to eliminate the remaining systematic errors 
which the PCA technique cannot remove.

\acknowledgements
We thank Gary Bernstein for collaborative work on this survey 
and many helpful discussions. 
We thank Eric Linder, Masahiro Takada, Ludovic van Waerbeke
and Martin White for helpful comments. This work is supported in 
part by NASA grant NAG5-10924 and by a Keck foundation grant.

\end{document}